\documentclass[11pt]{article}
\usepackage[dvips]{epsfig}

 \advance \textwidth by -2in
 \advance \textheight by -3in
\def\##1{{\bf{#1}}}
\def\=#1{\underline{\underline{#1}}}

\def\+#1{\underline{\bf #1}}
\def\*#1{\underline{\underline{\bf #1}}}

\def\.{\mbox{ \tiny{$^\bullet$} }}

\def\ux{\#{\hat u}_x}

\def\c#1{\cite{#1}}

\def\r#1{(\ref{#1})}

\def\epsr{\epsilon_r}
\def\mur{\mu_r}

\begin{document}

\noindent {\large {\bf SCATTERING BY A NIHILITY SPHERE}} \vskip 0.2cm

\noindent  {\bf Akhlesh Lakhtakia\footnote{E--mail: akhlesh@psu.edu. Also affiliated with: Department of
Physics, Imperial College, London SW7 2AZ, United Kingdom}
}
\vskip 0.2cm

\noindent {\sf CATMAS~---~Computational \& Theoretical Materials Sciences Group \\
\noindent Department of Engineering Science \& Mechanics\\
\noindent 212 Earth \& Engineering Sciences Building\\
\noindent Pennsylvania State University, University Park, PA
16802--6812}

\vskip 0.4cm

\noindent {\bf ABSTRACT:} On interrogation by a plane wave,
the back--scattering efficiency of
a nihility sphere is identically zero, and its extinction and 
forward--scattering efficiencies are higher than those of a perfectly conducting
sphere.

\vskip 0.2cm \noindent {\bf Keywords:} {\em   back--scattering efficiency; extinction
efficiency;
forward--scattering efficiency;
nihility }

\vskip 0.4cm

\section{INTRODUCTION}

The concept of nihility as an electromagnetic {\em medium} has emerged \c{ET} from
the rather extraordinary developments on negatively refracting materials during
this decade \c{LMW_aeu, Rama}. 
Whereas the relative permittivity and relative permeability of vacuum
are $\epsr=\mur=1$, and those of anti--vacuum are $\epsr=\mur=-1$,
those of nihility are $\epsr=\mur=0$. The so--called perfect lens
of Pendry \c{Pen} is made of anti--vacuum \c{PLN1}, and any perfect
lens in the present context is required to simulate nihility \c{PLN1,PLN2}.

Although quite some attention has been devoted to the electromagnetic response
characteristics of anti--vacuum (or some approximation thereof),
nihility has been neglected in comparison \c{TNSMS,Ziol}. Reflection
and refraction of plane waves due to nihility half--spaces has recently been reported
\c{LM05}, and so has the plane--wave response of nihility slabs \c{PLN1,PLN2}. 
Along the same lines, this communication focuses on the plane--wave
response of a nihility sphere. An $\exp(-i\omega t)$ time--dependence is implicit
in the following sections.

\section{THEORY}
Consider the spherical region $r<a$ occupied by nihility, whereas the region
$r>a$ is vacuous. Without loss of generality, the incident plane wave is taken
to be linearly polarized and traveling along the $+z$ axis; thus,
\begin{equation}
\#E_{inc}(\#r) = E_0\, \ux\,\exp(ik_0z)\,,
\end{equation}
where $E_0$ is the amplitude, $k_0$ is the free--space wavenumber, and $\ux$ is
the unit vector parallel to the $+x$ axis. 

As is commonplace, the incident plane wave
is represented in terms of vector spherical harmonics
$\#M_{\sigma mn}^{(j)}(\#w)$ and $\#N_{\sigma mn}^{(j)}(\#w)$ \c{Str,BH} as
follows:
\begin{equation}
\#E_{inc}(\#r) =E_0 \sum_{n=1}^\infty\,i^n\frac{2n+1}{n(n+1)}\,
\left[\#M_{o1n}^{(1)}(k_0\#r) -i\#N_{e1n}^{(1)}(k_0\#r)\right]\,.
\end{equation}
The scattered field is also stated in terms of vector spherical harmonics
as \c{BH}
\begin{equation}
\#E_{sc}(\#r) =E_0 \sum_{n=1}^\infty\,i^n\frac{2n+1}{n(n+1)}\,
\left[ia_n\#N_{e1n}^{(3)}(k_0\#r) -b_n\#M_{o1n}^{(3)}(k_0\#r)\right]\,,
\quad r \geq a\,,
\end{equation}
where
\begin{equation}
a_n= \frac{ \epsr\, j_n(N\xi)\, \psi_n^{(1)}(\xi) -  j_n(\xi)\,\psi_n^{(1)}(N\xi)}
{ \epsr\, j_n(N\xi)\, \psi_n^{(3)}(\xi) - h_n^{(1)}(\xi)\,\psi_n^{(1)}(N\xi)}\,
\label{an}
\end{equation}
and
\begin{equation}
b_n=\frac{ \mur\, j_n(N\xi) \,\psi_n^{(1)}(\xi) -   j_n(\xi)\,\psi_n^{(1)}(N\xi)}
{ \mur\, j_n(N\xi) \,\psi_n^{(3)}(\xi) -  h_n^{(1)}(\xi)\,\psi_n^{(1)}(N\xi)}\,;
\label{bn}
\end{equation}
$j_n(\xi)$ and $h_n^{(1)}(\xi)$ are the spherical Bessel function
and the spherical Hankel function of the first kind; 
\begin{equation}
\psi_n^{(1)}(w) = \frac{d}{dw}\left[w\,j_n(w)\right]\,
\end{equation}
and
\begin{equation}
\psi_n^{(3)}(w) = \frac{d}{dw}\left[w\,h_n^{(1)}(w)\right]\,;
\end{equation}
$N=\sqrt{\epsr\mur}$
and $\xi=k_0a$; and $\epsr$ and $\mur$ are, respectively, the relative permittivity
and the relative permeability of the matter occupying the region $r< a$.

Taking the limits $\epsr\to0$ and $\mur\to0$ for the scattering medium (i.e.,
nihility), we obtain
\begin{equation}
a_n=b_n=\frac{j_n(\xi)}{h_n^{(1)}(\xi)}\,.
\label{anbn}
\end{equation}
The equality of the coefficients $a_n=b_n\, \forall n $ is remarkable,
and possibly unique to nihility spheres.

\section{DISCUSSION}

\begin{figure}[!ht]
\centering \psfull \epsfig{file=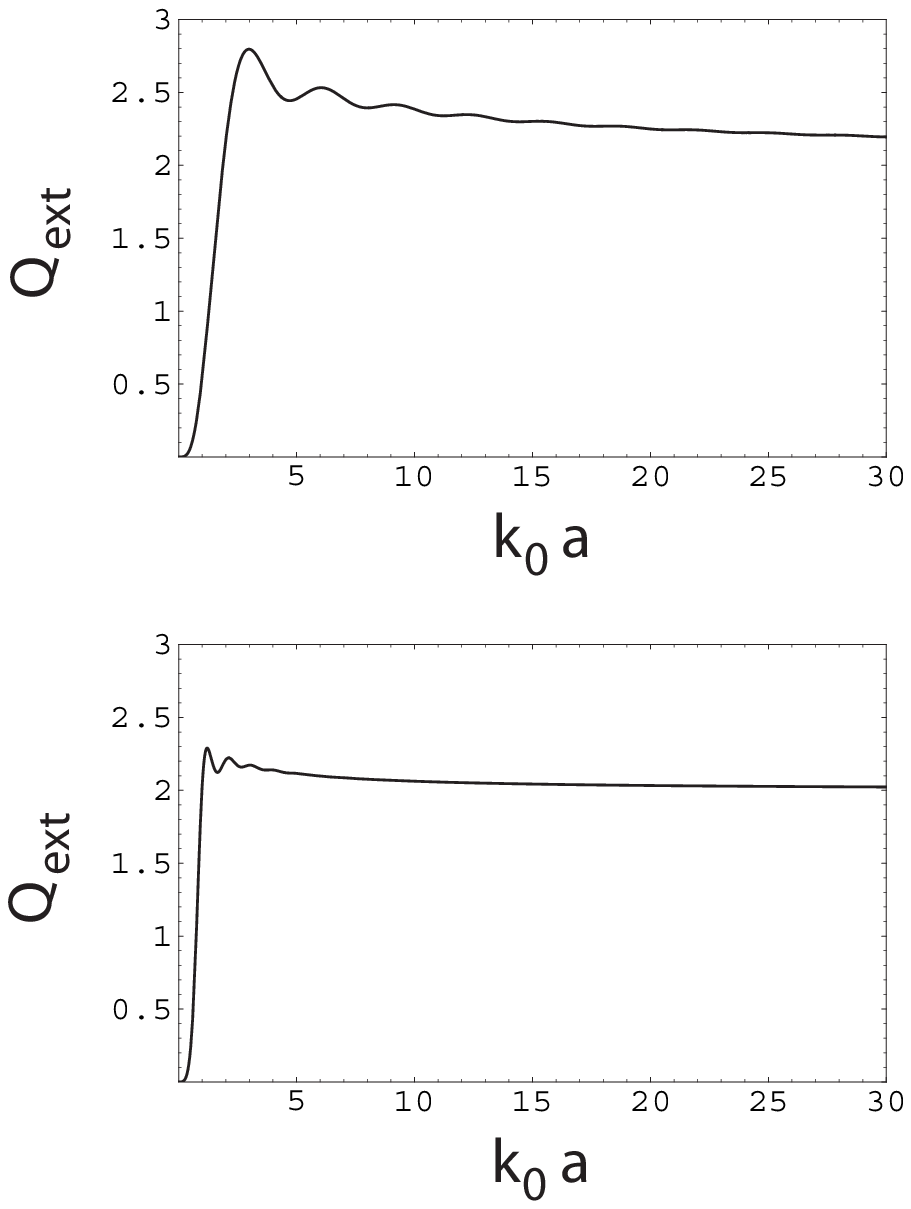,width=3.0in}
 \caption{\label{Fig1}  Extinction efficiency $Q_{ext}$ of a sphere
 as a function of its normalized radius $\xi=k_0a$. Top: nihility sphere. Bottom:
 perfectly conducting sphere.
  }
\end{figure}

Figure \ref{Fig1} contains a plot of the extinction efficiency 
\begin{equation}
Q_{ext}=\frac{2}{\xi^2}\,\sum_{n=1}^\infty\,
(2n+1)\, \Re\left(a_n+b_n\right) 
\end{equation}
of the nihility sphere
as a function of its normalized radius $\xi$. The overall profile is similar to that 
for a perfectly conducting sphere \c{VLV}, which is also shown in the same figure,
but extinction by the nihility sphere is larger than that by a perfectly
conducting sphere. Furthermore, the peak extinction by a nihility sphere
occurs at a larger value of $\xi$ ($\approx2.981$) than by a perfectly
conducting sphere ($\xi\approx1.209$).

Calculations associated with \r{an} and \r{bn}
show that no electromagnetic field exists inside nihility spheres. Hence,
there is no absorption, and
the extinction efficiency equals the total scattering efficiency \c{BH}
\begin{equation}
Q_{sca}=\frac{2}{\xi^2}\,\sum_{n=1}^\infty\,
(2n+1)\,  \left(\vert a_n\vert^2+\vert b_n\vert^2\right) \,.
\end{equation}

\begin{figure}[!ht]
\centering \psfull \epsfig{file=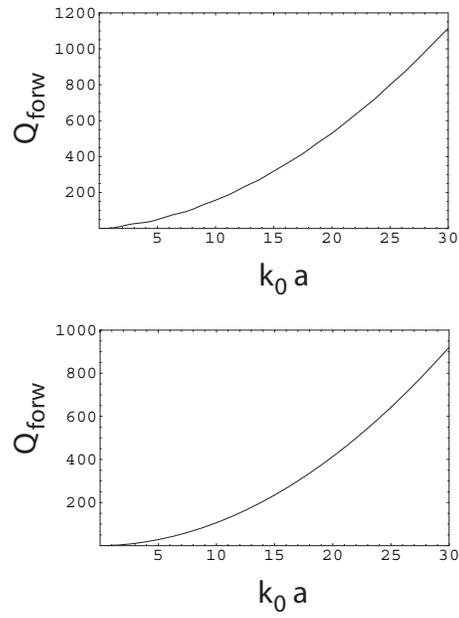,width=3.0in}
 \caption{\label{Fig2}  Forward--scattering efficiency $Q_{forw}$ of a sphere
 as a function of its normalized radius $\xi=k_0a$. Top: nihility sphere. Bottom:
 perfectly conducting sphere.
  }
\end{figure}

The forward--scattering efficiency  
\begin{equation}
Q_{forw}=\frac{1}{\xi^2}\Big\vert\sum_{n=1}^\infty\,
(2n+1)\,  \left(a_n+ b_n\right) \Big\vert^2\,
\end{equation}
of a nihility sphere is plotted in Figure \ref{Fig2} as a function
of $\xi$, and compared with that of a perfectly conducting sphere.
That of the nihility sphere is higher.

The most interesting feature of the plane--wave response of
a nihility sphere is its back--scattering efficiency \c{VB}
\begin{equation}
Q_{back}=\frac{1}{\xi^2}\Big\vert\sum_{n=1}^\infty\,(-)^n\,
(2n+1)\,  \left(b_n- a_n\right) \Big\vert^2\,.
\end{equation}
By virtue of \r{anbn}, 
\begin{equation}
Q_{back}\equiv 0
\label{Qb}
\end{equation}
 for a nihility
sphere; of course, $Q_{back}\ne 0$ for perfectly conducting spheres \c{VB,VLV}.

Equation \r{Qb} is a remarkable result, because it implies that the probability of detection
of a nihility sphere by a monostatic radar system is very low. This result would not change even if the ambient (isotropic) medium were to have relative permittivity and relative permeability other than unity. This conclusion can be understood by realizing  that nihility is impedance--matched to any isotropic, achiral, dielectric--magnetic medium
\c{Usl}.

\newpage

\newpage

\end{document}